\documentstyle[a4wide,12pt]{article}
\title{A metric of Yukawa potential as an exact solution\\
to the field equations of general relativity}
\author{Vu B Ho\\Department of Physics\\Monash University\\
Clayton Victoria 3168\\Australia}
\date{}
\begin{document}
\maketitle
\begin{abstract}
It is shown that, by defining a suitable energy momentum tensor, the field
equations of general relativity admit a line element of Yukawa potential as
an exact solution. It is also shown that matter that produces strong force may
be negative, in which case there would be no Schwarzschild-like singularity.
\end{abstract}
\newpage

\noindent
{\bf 1. Approximate solution}\\
\\
It has been shown that the line element of Yukawa potential of the form
\begin{equation}
e^{-\nu}=1-\alpha\frac{e^{-\beta r}}{r}
\end{equation}
approximately satisfies, at short range $\beta r\ll 1$, the vacuum field
equations of general relativity, $T_{\mu\nu}=0$,
\begin{equation}
R_{\mu\nu}-\frac{1}{2}g_{\mu\nu}R=\kappa T_{\mu\nu},
\end{equation}
assuming a centrally symmetric spacetime metric
\begin{equation}
ds^2=e^\mu dt^2-e^\nu dr^2-r^2(d\theta^2+sin^2\theta d\phi^2).
\end{equation}

In general, the quantity $R=1/\beta$ specifies a range so that the line
element of Yukawa potential can be approximated as a solution to the field
equations of general relativity for the region $r\ll R$. It is seen that
for the maximal possible range of $R\rightarrow\infty$, the metric of Yukawa
form reduces to the familiar Schwarzschild metric, which is
used to describe the gravitational field. In the case of short range of
nuclear physics, the quantity $R$ can be assigned a value in terms of the
fundamental constants $\hbar$ and $c$, and the rest mass of Yukawa force
carrier. The result has shown that within the short range of strong force,
the field equations of general relativity admit a line element that takes
the form of Yukawa potential for strong interaction. This leads to the
conclusion that if there is no other form of matter, besides the mass and
the charge, that characterises strong interaction, then it would be
possible to consider strong interaction also a manifestation of general
relativity at short range.

Assume the motion of a particle in a strong force field is governed by the
geodesic equations
\begin{equation}
\frac{d^2x^\mu}{ds^2}+\Gamma^\mu_{\nu\sigma}\frac{dx^\nu}{ds}
\frac{dx^\sigma}{ds}=0,
\end{equation}
then with the metric of Yukawa potential of the form
\begin{equation}
ds^2=\left(1-\alpha\frac{e^{-\beta r}}{r}\right)c^2dt^2 - \left(1-
\alpha\frac{e^{-\beta r}}{r}\right)^{-1}dr^2 -
r^2\left(d\theta^2+sin^2\theta d\phi^2\right),
\end{equation}
the equations for the geodesics can be written explicitly as \cite{Berg,Lawd}
\begin{equation}
\left(1-\alpha\frac{e^{-\beta r}}{r}\right)^{-1}\left(\frac{dr}{d\tau}
\right)^2 + r^2\left(\frac{d\theta}{d\tau}\right)^2 +
r^2sin^2\theta\left(\frac{d\phi}{d\tau}\right)^2 - c^2\left(1-\frac{\alpha
e^{-\beta r}}{r}\right)\left(\frac{dt}{d\tau}\right)^2=-c^2,
\end{equation}
\begin{eqnarray}
\frac{d}{d\tau}\left(r^2\frac{d\theta}{d\tau}\right) - r^2\sin\theta
\cos\theta\left(\frac{d\phi}{d\tau}\right)^2&=&0,\\
\frac{d}{d\tau}\left(r^2\sin^2\theta\frac{d\phi}{d\tau}\right)&=&0,\\
\frac{d}{d\tau}\left[\left(1-\alpha\frac{e^{-\beta r}}{r}\right)
\frac{dt}{d\tau}\right]&=&0.
\end{eqnarray}
By choosing spherical polar coordinates and considering the motion in the
plane $\theta=\pi/2$, the equations (8) and (9) are reduced to
\begin{equation}
\frac{d\phi}{d\tau}=\frac{l}{r^2}, \ \ \ \ \ \ \
\frac{dt}{d\tau}=\frac{kr}{r-\alpha e^{-\beta r}},
\end{equation}
where $l$ and $k$ are constants of integration. With these relations, the
equation for the orbit can be obtained from the equation (6) as
\begin{equation}
\left(\frac{l}{r^2}\frac{dr}{d\phi}\right)^2 + \frac{l^2}{r^2} =
c^2(k^2-1) + \frac{\alpha c^2}{r}e^{-\beta r} + \frac{\alpha l^2}{r^3}e^{-
\beta r}.
\end{equation}
In the case when the condition $\beta r\ll 1$ is satisfied, with the
approximation $e^{-\beta r}=1-\beta r$, the equation for the orbit becomes
\begin{equation}
\left(\frac{l}{r^2}\frac{dr}{d\phi}\right)^2 +
\frac{l^2(1+\alpha\beta)}{r^2} = c^2(k^2-(1+\alpha\beta)) + \frac{\alpha
c^2}{r} + \frac{\alpha l^2}{r^3}.
\end{equation}
By letting $u=1/r$ and differentiating the resulting equation with respect
to the variable $\phi$, it is found
\begin{equation}
\frac{d^2u}{d\phi^2}+(1+\alpha\beta)u = \frac{\alpha c^2}{2l^2} +
\frac{3\alpha l^2}{2}u^2.
\end{equation}
Hence, the dynamics of a particle under the influence of strong force of
Yukawa potential is similar to that of a particle in the Schwarzschild
gravitational field.\\
\\
\noindent
{\bf 2. Exact solution}\\
\\
Unlike the assumed massless force carriers of the gravitational field, hence
the vacuum solution to the field equations of general relativity could be
justified, the force carriers of the strong field are assumed to be
massive. Therefore, any solutions to the field equations of general
relativity that may be used to describe the strong field should be a
nonvacuum solution. This means that in order to find a more appropriate
solution to describe strong force, a strong energy momemtum tensor must be
specified. In the following it will be discussed a particular form of
strong energy momentum tensor so that the field equations of general
relativity will admit the line element of Yukawa potential as an exact
solution.

Consider a strong energy momentum tensor of the form\\
\\
\begin{equation}
T_\mu^\nu = \left(\begin{array}{cccc}
-\frac{\alpha\beta}{\kappa}\frac{e^{-\beta r}}{r^2} & 0 & 0 & 0\\
0 & -\frac{\alpha\beta}{\kappa}\frac{e^{-\beta r}}{r^2} & 0 & 0\\
0 & 0 & \frac{\alpha\beta^2}{2\kappa}\frac{e^{-\beta r}}{r} & 0\\
0 & 0 & 0 & \frac{\alpha\beta^2}{2\kappa}\frac{e^{-\beta r}}{r}
\end{array}\right)
\end{equation}
\\
With this energy momentum tensor, the field equations of general relativity
reduce to the system of equations \cite{Land}
\begin{eqnarray}
e^{-\nu}\left(\frac{\partial\nu}{\partial r}-\frac{1}{r}\right)+\frac{1}{r}
&=&-\alpha\beta\frac{e^{-\beta r}}{r},\\
-e^{-\nu}\left(\frac{\partial\mu}{\partial r}+\frac{1}{r}\right)+\frac{1}{r}
&=&-\alpha\beta\frac{e^{-\beta r}}{r},\\
\frac{\partial\nu}{\partial t}&=&0,
\end{eqnarray}
\begin{equation}
-e^{-\nu}\left(2\frac{\partial^2\mu}{\partial r^2}+\left(\frac{\partial\mu}
{\partial r}\right)^2+\frac{2}{r}\left(\frac{\partial\mu}{\partial r}-
\frac{\partial\nu}{\partial r}\right)-
\frac{\partial\mu}{\partial r}\frac{\partial\nu}{\partial r}\right) +
e^{-\mu}\left(2\frac{\partial^2\nu}{\partial t^2}+\left(\frac{\partial\nu}
{\partial t}\right)^2-\frac{\partial\nu}{\partial t}\frac{\partial\mu}
{\partial t}\right)=\alpha\beta^2\frac{e^{-\beta r}}{r}.
\end{equation}

These equations are not independent since it can be verified that the last
equation follows from the first three equations. Furthermore, the first two
equations give $\partial\nu/\partial r+\partial\mu/\partial r=0$ that also
leads to $\nu+\mu=0$. This system of equations when integrated gives the
metric of Yukawa potential. Actually, the more complete solution of the above
system of equations is of the form
\begin{equation}
e^{-\nu}=1-\alpha\frac{e^{-\beta r}}{r}+\frac{c}{r}.
\end{equation}
The term $c/r$ may account for the Coulomb repulsive force between two
charged particles such as two protons. In this case the quantity $\alpha$ may
be determined through scattering processes or $\alpha$-decays. However, if it
is considered only strong forces then this term may be ignored by letting
$c=0$.

The form of the energy momentum tensor considered above is mathematically
acceptable since it can be verified to satisfy the conservation laws
\begin{equation}
T^\nu_{\mu;\nu}=\frac{1}{\sqrt{-g}}\frac{\partial T^\nu_\mu\sqrt{-g}}
{\partial x^\nu} - \frac{1}{2}\frac{g_{\lambda\sigma}}{\partial x^\mu}
T^{\lambda\sigma}=0,
\end{equation}
using the metric $g_{\mu\nu}$ of Yukawa form.

However, there emerges an important feature that relates to the nature of
the quantity $\alpha$ in the line element and that of the energy momentum
tensor. That is, since the quantities $\kappa$ and $\beta$ are positive,
the energy component $T_0^0$ and the quantity $\alpha$ always have opposite
signs. Therefore, since $g^{00}$ is positive,  if the energy component
$T_{00}$ is considered to be positive, then the energy component
$T^0_0=g^{00}T_{00}$ must be positive, and, in this case, the quantity
$\alpha$ must be negative and there would be no Schwarzschild-like singularity.

\end{document}